# Cuprate High Temperature Superconductors

**Roland Hott & Thomas Wolf**
Karlsruher Institut für Technologie, Institut für Festkörperphysik,
P. O. Box 3640, 76021 Karlsruhe, GERMANY

## 1. INTRODUCTION

Cuprate High-Temperature Superconductors (“HTS”) have played an outstanding role in the scientific and technological development of superconductors. Except for semiconductors, no other class of materials has been investigated so thoroughly by thousands and thousands of researchers worldwide. The plethora of preparational degrees of freedom, the inherent tendency towards inhomogeneities and defects did not allow easy progress in the preparation of these materials. The very short SC coherence lengths of the order of the dimensions of the crystallographic unit cell were on one hand a high hurdle. On the other hand, this SC link of nano-scale microstructure and macroscopic transport properties provided minute monitoring of remnant obstructive defects which could then be tackled by further materials optimization. For cuprate HTS, many material problems have been solved or at least thoroughly discussed which had not been even realized as a problem for other superconductors before.

## 2. STRUCTURAL ASPECTS

The structural element of HTS compounds related to the location of mobile charge carriers are stacks of a certain number n = 1, 2, 3, ... of $CuO_2$ layers "glued" on top of each other by means of intermediate Ca layers (see Fig. 1b) [2,3,4,5]. Counterpart of these "*active blocks*" of $(CuO_2/Ca/)_{n-1}CuO_2$ stacks are "*charge reservoir blocks*" $EO/(AO_x)_m/EO$ with m = 1, 2 monolayers of a quite arbitrary oxide $AO_x$ (A = Bi [3], Pb [4], Tl [3], Hg [3], Au [6], Cu [3], Ca [7], B [4], Al [4], Ga [4]; see Table 1) "wrapped" on each side by a monolayer of alkaline earth oxide EO with E = Ba, Sr (see Fig. 1b). The HTS structure results from alternating stacking of these two block units. The choice of BaO or SrO as "wrapping" layer is not arbitrary but depends on the involved $AO_x$ since it has to provide a good spatial adjustment of the $CuO_2$ to the $AO_x$ layers.

The general chemical formula $\mathbf{A_m}E_\mathbf{2}Ca_\mathbf{n-1}Cu_\mathbf{n}O_{2n+m+2+y}$ (see Fig. 1b) is conveniently abbreviated as **A-m2(n-1)n** [5] (e. g. $Bi_2Sr_2Ca_2Cu_3O_{10}$: Bi-2223) neglecting the indication of the alkaline earth element (see Tab.1). The family of all n = 1, 2, 3, ... representatives with common $AO_x$ are often referred to as “A-HTS”, e. g. Bi-HTS. The most prominent compound $\mathbf{YB}a_2\mathbf{C}u_3\mathbf{O}_7$ (see Fig. 1a), the first HTS discovered with a critical temperature $T_c$ above the boiling point of liquid nitrogen [8], is traditionally abbreviated as "**YBCO**" or "**Y-123**" ($\mathbf{Y_1}Ba_\mathbf{2}Cu_\mathbf{3}O_{7-\delta}$). It also fits into the general HTS classification scheme as a modification of Cu-1212 where Ca is completely substituted by Y. This substitution introduces extra negative charge in the $CuO_2$ layers due to the higher valence of Y (+3) compared to Ca (+2).

The HTS compounds $REBa_2Cu_3O_{7-\delta}$ ("RBCO", "RE-123") where RE can be La [3] or any other rare earth element [9] except for Ce or Tb [10] can be regarded as a generalization of this substitution scheme. The lanthanide contraction of the RE ions provides here an experimental handle on the distance between the two $CuO_2$ layers of the active block of the doped Cu-1212 compound [11,12,13]. $\mathbf{Y_1}Ba_\mathbf{2}Cu_\mathbf{4}O_8$ ("**Y-124**") [14] is the m = 2 counterpart Cu-2212 of YBCO.

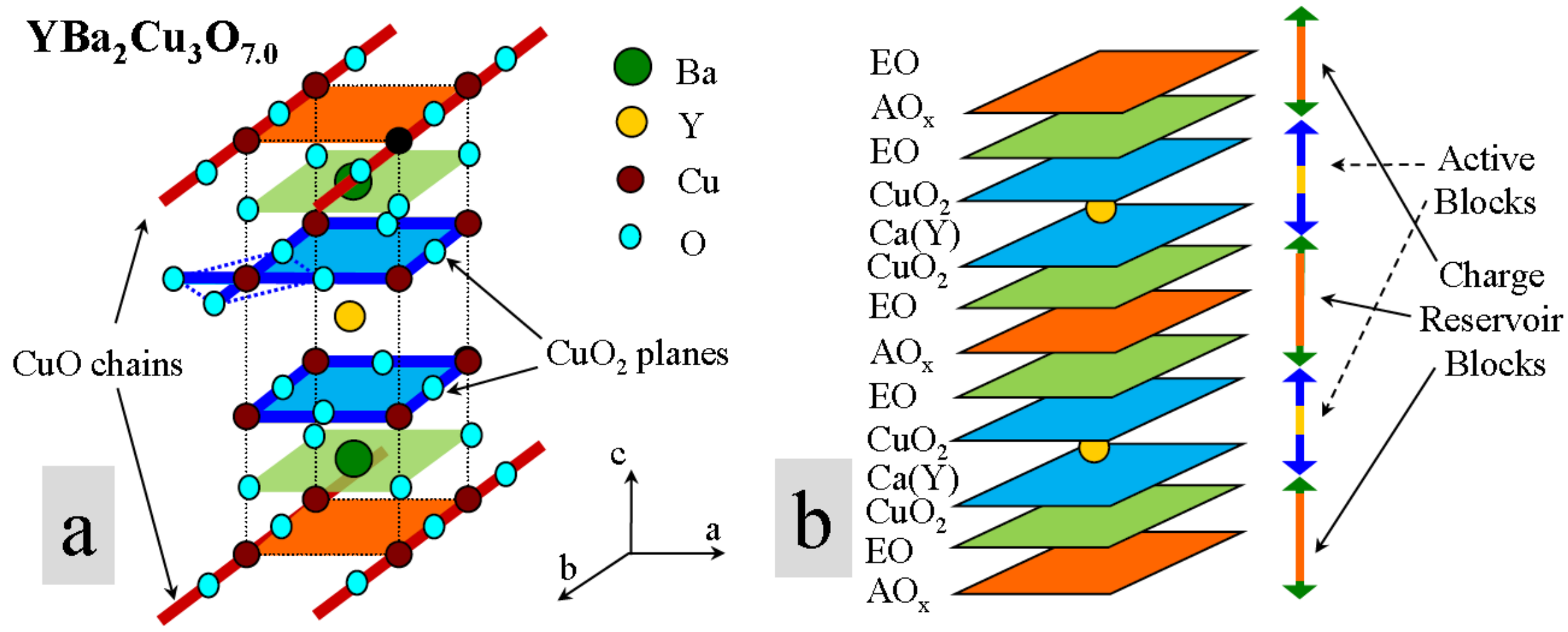

**Fig. 1** a) Crystal structure of $YBa_2Cu_3O_7$ ("YBCO"). The presence of the CuO chains introduces an orthorhombic distortion of the unit cell (a = 0.382 nm, b = 0.389 nm, c = 1.167 nm [1]).
b) General structure of a cuprate HTS A-m2(n-1)n ($A_mE_2Ca_{n-1}Cu_nO_{2n+m+2+y}$) for m = 1. For m = 0 or m = 2 the missing (additional) $AO_x$ layer per unit cell leads to a (a/2, b/2, 0) "side step" of the unit cells adjoining in c-axis direction.

The "**214**" HTS compounds $E_2Cu_1O_4$ (see Tab. 1), e. g. $La_{2-x}Sr_xCuO_4$ ("LSCO") or $Nd_{2-x}Ce_xCuO_4$ ("NCCO") are a bit exotic in this ordering scheme but may also be represented here as “0201” with m = 0, n = 1 and $E_2 = La_{2-x}Sr_x$ and $E_2 = Nd_{2-x}Ce_x$, respectively.

Further interesting chemical modifications of the basic HTS compositions are the introduction of fluorine [40,41] or chlorine [42] as more electronegative substituents of oxygen. For Hg-1223, $T_c$ = 135 K can thus be raised to 138 K [43], the highest $T_c$ reported by now under normal pressure conditions (164 K at 30 GPa [44]).

## 3. METALLURGICAL ASPECTS

Within 27 years since their discovery [45], cuprate HTS samples have greatly improved towards high materials quality and can in fact nowadays be prepared in a remarkably reproducible way. The enormous worldwide efforts for this achievement may be estimated from the plain number of more than 100 000 articles which meanwhile have been published on cuprate high-$T_c$ superconductivity, strongly outnumbering the only about 15 000 publications [46,47,48] that appeared within a whole century on the remaining superconducting materials.

Nevertheless, the HTS materials quality level is still far from the standards of classical superconductors. The reason is the larger number of at least four chemical elements from which the various cuprate HTS phases have to be formed: The majority of the classical SC compounds is made of only two elements. Each element contributes an additional degree of freedom to the preparation route towards new compounds. For the metallurgist, this translates roughly into one order of magnitude more elaborate exploratory efforts. A new ternary cuprate

HTS compound requires therefore typically as much metallurgical optimization work as 100 binary compounds.

With respect to the cation stoichiometry, for the RE-123 phase as a particularly well-investigated HTS example the RE/Ba ratio represents the second chemical degree of freedom which may also deviate from the stoichiometric value 1/2. Y-123 (YBCO) represents here a remarkable exception. However, RE-123 compounds where the Y ions are replaced by larger RE ions like Gd or Nd exhibit a pronounced homogeneity range. For Nd-123, the Nd/Ba ratio encountered in such solid solutions may vary from 0.49 to 2, with the oxygen content being still an almost fully independent chemical variable. The Cu/(EA=Ba or Sr) ratio as chemical degree of freedom of all “hole-doped” HTS compounds (see Tab. 1) has up to now not been studied in great detail. For the 123 phase as a particularly complicated HTS example with respect to the Cu content featuring two $CuO_2$ layers and an additional CuO chain structure in the unit cell, the present assumption is that the Cu/Ba ratio sticks to the stoichiometric value 3/2. For samples prepared close to their peritectic temperatures or at very low temperatures close to the boundary of the stability field of the 123 phase this may no longer be the case.

Chemical purity of the starting material is still a topical issue for a reproducible preparation of cuprate HTS. The use of chemicals with a purity of 99.99% and better is mandatory but still not sufficient. The frequently used Ba source material $BaCO_3$ is usually not completely reacted and may thus lead to the incorporation of carbonate ions into the HTS cuprate phase. During the preparation procedure, the formation of even a small amount of liquid has to be carefully avoided since this may corrode the substrate or crucible or may solidify at grain boundaries (see Fig. 5) thus impeding the grain-to-grain supercurrent transfer. This corrosion process is particular harmful for crystal growth experiments where the complete melt encounters the crucible wall. BaO/CuO melts are highly corrosive and attack all conventional types of crucible materials forming new solid phases. Some of these reaction products have turned out to be well suited as materials for corrosion resistant crucibles for the preparation of cuprate HTS. $BaZrO_3$ is here the best-known example [29], $BaHfO_3$ and $BaSnO_3$ are promissing candidates.

The preparation of an HTS layer structure based on $(CuO_2/Ca/)_{n-1}CuO_2$ stacks with a well-defined number $n$ of $CuO_2$ layers (see Fig. 1) introduces another challenge: As the formation enthalpy of a compound, e. g. with a single $CuO_2$ layer ($n$ = 1), differs only little from that of the ($n$ = 2)-compound with two adjacent $CuO_2$ layers, these materials tend to form polytypes [14].

## 4. STRUCTURE AND $T_C$

Experimentally, for all HTS families A-m2(n-1)n the optimized $T_c$ is found to increase from n = 1 to n = 3, 4 and to decrease again for higher n (s. Tab.1). It is still unclear whether this $T_c$ maximum is an intrinsic HTS property since the synthesis of higher-n members of the HTS families turns out to be more and more complicated [49,50,51,52]. In particular, there is at present no preparation technique that allows to adjust here a sufficiently high oxygen content or to provide otherwise sufficient electronic doping that would allow to clarify the possible range of $T_c$ optimization for higher n HTS A-m2(n-1)n [49,52]. Hence the question is still open if such an optimized $T_c$ may eventually continue to increase towards higher n, possibly up to $T_c$ ~ 200 K [53].

| HTS Family | Stochiometry | Notation | Compounds | Highest $T_c$ |
|---|---|---|---|---|
| Bi-HTS | $Bi_mSr_2Ca_{n-1}Cu_nO_{2n+m+2}$ | | Bi-1212 | 102 K [15] |
| | | Bi-m2(n-1)n, | Bi-2201 | 34 K [16] |
| | m = 1, 2 | BSCCO | Bi-2212 | 96 K [17] |
| | n = 1, 2, 3 . . . | | Bi-2223 | 110 K [3] |
| | | | Bi-2234 | 110 K [18] |
| Pb-HTS | $Pb_mSr_2Ca_{n-1}Cu_nO_{2n+m+2}$ | Pb-m2(n-1)n | Pb-1212 | 70 K [19] |
| | | | Pb-1223 | 122 K [20] |
| Tl-HTS | $Tl_mBa_2Ca_{n-1}Cu_nO_{2n+m+2}$ | | Tl-1201 | 50 K [3] |
| | m = 1, 2 | Tl-m2(n-1)n, | Tl-1212 | 82 K [3] |
| | n = 1, 2, 3 . . . | TBCCO | Tl-1223 | 133 K [21] |
| | | | Tl-1234 | 127 K [22] |
| | | | Tl-2201 | 90 K [3] |
| | | | Tl-2212 | 110 K [3] |
| | | | Tl-2223 | 128 K [23] |
| | | | Tl-2234 | 119 K [24] |
| Hg-HTS | $Hg_mBa_2Ca_{n-1}Cu_nO_{2n+m+2}$ | | Hg-1201 | 97 K [3] |
| | | Hg-m2(n-1)n, | Hg-1212 | 128 K [3] |
| | m = 1, 2 | HBCCO | Hg-1223 | 135 K [25] |
| | n = 1, 2, 3 . . . | | Hg-1234 | 127 K [25] |
| | | | Hg-1245 | 110 K [25] |
| | | | Hg-1256 | 107 K [25] |
| | | | Hg-2212 | 44 K [26] |
| | | | Hg-2223 | 45 K [27] |
| | | | Hg-2234 | 114 K [27] |
| Au-HTS | $Au_mBa_2Ca_{n-1}Cu_nO_{2n+m+2}$ | Au-m2(n-1)n | Au-1212 | 82 K [6] |
| 123-HTS | $REBa_2Cu_3O_{7-\delta}$ | RE-123, RBCO | Y-123, YBCO | 92 K [11] |
| | RE = Y, La, Pr, Nd, Sm, | | Nd-123, NBCO | 96 K [11] |
| | Eu, Gd, Tb, Dy, Ho, | | Gd-123 | 94 K [28] |
| | Er, Tm, Yb, Lu | | Er-123 | 92 K [**29**] |
| | | | Yb-123 | 89 K [9] |
| Cu-HTS | $Cu_mBa_2Ca_{n-1}Cu_nO_{2n+m+2}$ | Cu-m2(n-1)n | Cu-1223 | 60 K [3] |
| | | | Cu-1234 | 117 K [30] |
| | m = 1, 2 | | Cu-2223 | 67 K [3] |
| | n = 1, 2, 3 . . . | | Cu-2234 | 113 K [3] |
| | | | Cu-2245 | < 110 K [3] |
| Ru-HTS | $RuSr_2GdCu_2O_8$ | Ru-1212 | Ru-1212 | 72 K [31] |
| B-HTS | $B_mSr_2Ca_{n-1}Cu_nO_{2n+m+2}$ | B-m2(n-1)n | B-1223 | 75 K [32] |
| | | | B-1234 | 110 K [32] |
| | | | B-1245 | 85 K [32] |
| 214-HTS | $E_2CuO_4$ | LSCO | $La_{2-x}Sr_xCuO_4$ | 51 K [33] |
| | | "0201" | $Sr_2CuO_4$ | 25 (75)K [34] |
| | | *Electron-Doped HTS* | $La_{2-x}Ce_xCuO_4$ | 28 K [35] |
| | | PCCO | $Pr_{2-x}Ce_xCuO_4$ | 24 K [36] |
| | | NCCO | $Nd_{2-x}Ce_xCuO_4$ | 24 K [36] |
| | | | $Sm_{2-x}Ce_xCuO_4$ | 22 K [37] |
| | | | $Eu_{2-x}Ce_xCuO_4$ | 23 K [37] |
| | $Ba_2Ca_{n-1}Cu_nO_{2n+2}$ | "02(n-1)n" | "0212" | 90K [38] |
| | | | "0223" | 120K [38] |
| | | | "0234" | 105K [38] |
| | | | "0245" | 90K [38] |
| *Infinite-Layer* HTS | $ECuO_2$ | *Electron-Doped I. L.* | $Sr_{1-x}La_xCuO_2$ | 43 K [39] |

Table 1 Classification and reported optimized $T_c$ values of cuprate HTS compounds.

Another well-investigated experimental $T_c$ trend is the slight increase of the optimized $T_c$ values of the RE-123 HTS with increasing distance between the two $CuO_2$ layers in the active block. It has been explained in terms of a higher effective charge transfer to the $CuO_2$ layers [9,11]. For RE-123 HTS with larger RE ions (La, Pr, Nd,), sufficient oxygenation with respect to $T_c$ optimization becomes increasingly difficult. As a further complication these larger RE ions are comparable in size with the Ba ions which favors cation disorder with respect to the RE and Ba lattice sites [54] with the consequence of substantial $T_c$ degradation [55]. This disorder effect, oxygen deficiency and / or impurities had been suggested as reasons for the non-appearance or quick disappearance of superconductivity in Pr-123 [56] and Tb-123 as the only non-SC members of the 123-HTS family. Pr-123 samples with $T_c \sim 80$ K at normal pressure [57] and 105 K at 9.3 GPa [58] as measured immediately after preparation lost their SC properties within a few days. The nature of this SC Pr-123 phase is still unclear. It has been suggested that it consisted of Ba-rich Pr-123 which is unstable at room temperature with respect to spinodal decomposition.

In Bi-HTS, cation disorder at the Sr lattice site is inherent and strongly affects the value of $T_c$. In Bi-2201, partial substitution of Sr by RE = La, Pr, Nd, Sm, Eu, Gd, and Bi was shown to result in a monotonic decrease of $T_c$ with increasing ionic radius mismatch [17]. For Bi-2212, partial substitution of Ca by Y results in a $T_c$ optimum of 96 K at 8% Y doping, apparently due to a trade-off between the respective disorder effect and charge doping [17].

In 214-HTS without BaO or SrO "wrapping" layers around the $CuO_2$ layers (see Tab. 1), $T_c$ seems to be particularly sensitive with respect to oxygen disorder [59]. The electron doped 214-HTS such as $Nd_{2-x}Ce_xCuO_4$ ("NCCO") have here the additional complication of a different oxygen sublattice where oxygen ions on interstitial lattice positions in the $Nd_{2-x}Ce_xO_2$ layer (which are yet for hole doped 214-HTS the regular oxygen positions in the non-$CuO_2$ layer!) tend to suppress $T_c$ [60].

With respect to the $T_c$ dependence of the A-m2(n-1)n HTS families on the cation A of the charge reservoir blocks, there is an increase moving in the periodic table from Bi to Hg (see Tab. 1). However, continuing to Au, the reported $T_c$ is already substantially lower. This $T_c$ trend seems to be related to the chemical nature of the A-O bonds in the $AO_x$ layers [26].

For constant doping, buckling of the $CuO_2$ layer is observed to decrease $T_c$ [61]. Such deviations from the simple tetragonal crystal structure are found in most of the HTS compounds as a chemical consequence of the enforced $AO_x$ layer arrangement in the cuprate HTS structure. The record $T_c$ values for single $CuO_2$ layer HTS compounds of Tl-2201 ($T_c$ = 90 K)[62,63] and Hg-1201 ($T_c$ = 90 K)[64] with simple tetragonal crystal structure indicate that undistorted flat $CuO_2$ layers provide optimum superconductivity.

The SC critical temperature values $T_c$ reported in the preceeding paragraphs refer to the maximum values obtained for individually optimized doping either by variation of the oxygen content or by suitable substitution of cations. The following scenario applies to *hole*- [65] as well as to *electron-doping* of all cuprate HTS (see Fig. 2) [66]: The undoped compounds are antiferromagnetic insulators up to a critical temperature $T_N$ well above 300 K, with alternating spin orientations of the hole states that are localized around the Cu atoms in the $CuO_2$ layers. Adding charge carriers by doping relaxes the restrictions of spin alignment due to the interaction of these additional spin-1/2-particles with the spin lattice. $T_N$ decreases and the insulator turns into a "bad metal". At low temperature, however, the electric transport shows a dramatic change within a small doping range from an insulating to a SC behavior [70]. For $La_{2-x}Sr_xCuO_4$ this happens at a critical hole concentration x = 0.05 in the $CuO_2$ planes (see Fig. 1). On stronger doping, superconductivity can be observed up to an increasingly higher critical temperature $T_c$ until the maximum $T_c$ is achieved for "optimal doping" (x ≈ 0.16 for $La_{2-x}Sr_xCuO_4$). On further doping, $T_c$ decreases again until finally (x ≥ 0.27 for $La_{2-x}Sr_xCuO_4$) only normal conducting behavior is observed.

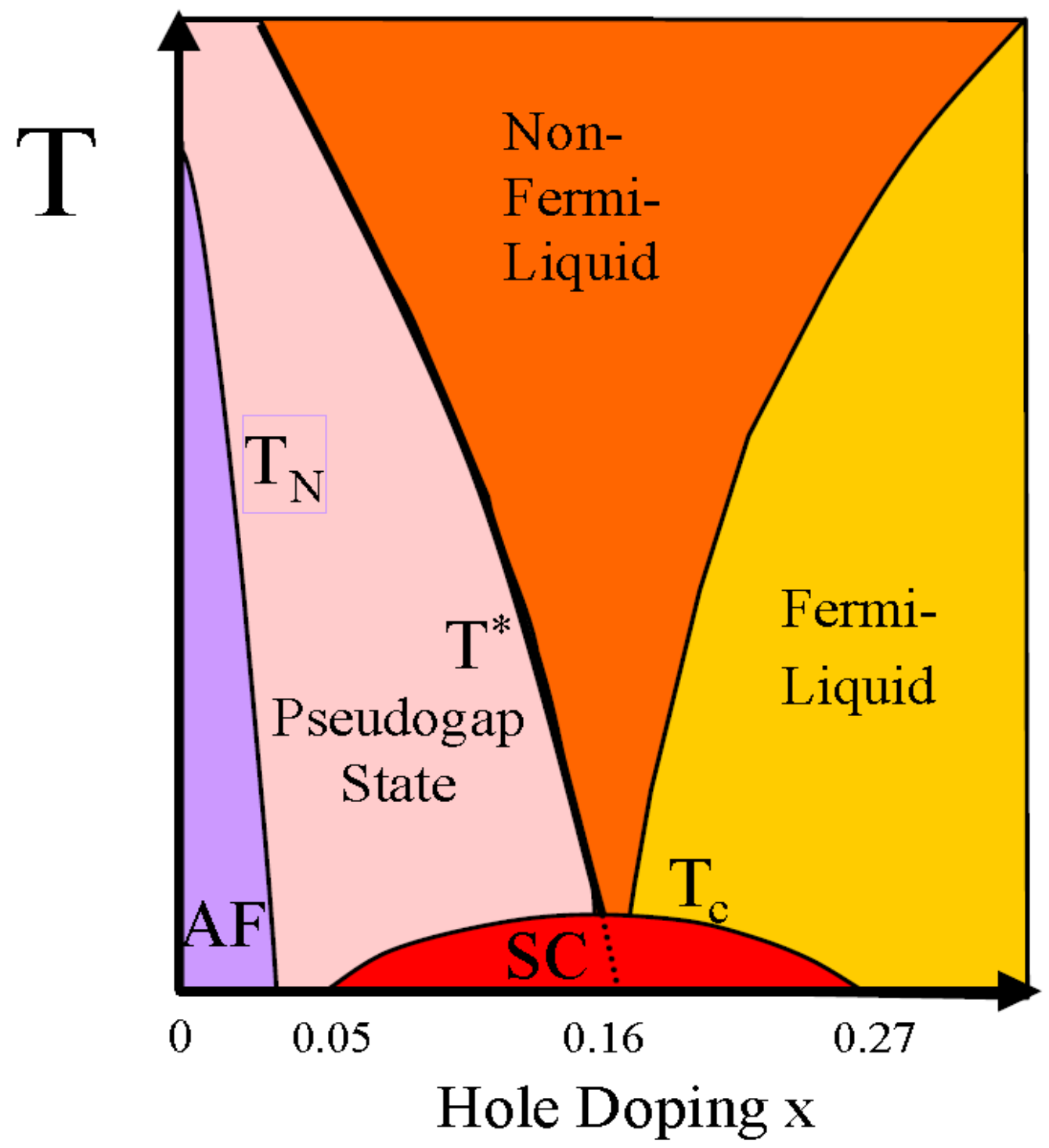

**Fig. 2** Schematic HTS temperature-doping phase diagram with the interplay of antiferromagnetism ("AF") and superconductivity ("SC") [65,67,68,69].

## 5. SUPERCONDUCTIVE COUPLING

The rationale that the phenomenon of superconductivity in HTS can be conceptually reduced to the physics of the $CuO_2$ layers [71] has evolved to a more and more 2-dimensional view in terms of $CuO_2$ *planes*. The superconductive coupling between these planes within a given $(CuO_2/Ca/)_{n-1}CuO_2$ stack ("interplane coupling") is much weaker than the intraplane coupling, but still much stronger than the superconductive coupling between the $(CuO_2/Ca/)_{n-1}CuO_2$ stacks which can be described as Josephson coupling (see Fig. 3).

The charge reservoir blocks $EO/(AO_x)_m/EO$ play in this idealized theoretical picture only a passive role providing the doping charge as well as the "storage space" for extra oxygen ions and cations introduced by additional doping. However, the huge pressure dependence of $T_c$ [44] in combination with the large quantitative variation of this effect for the various A-m2(n-1)n HTS families points to a more active role where the cations change not only their valency but also their transmission behavior for the interstack tunneling of Cooper pairs [72].

This becomes most evident for the Cu-HTS family, in particular for YBCO or the RE-123 HTS where the $AO_x$ layer is formed by 1-d CuO chain structures (see Fig. 1). There is experimental evidence that these CuO chains become SC, probably via proximity effect. The intercalation of superconductive CuO chain layers in-between the $CuO_2/Ca/CuO_2$ bilayer stacks is most likely the origin of the strong Josephson coupling between these bilayer stacks.

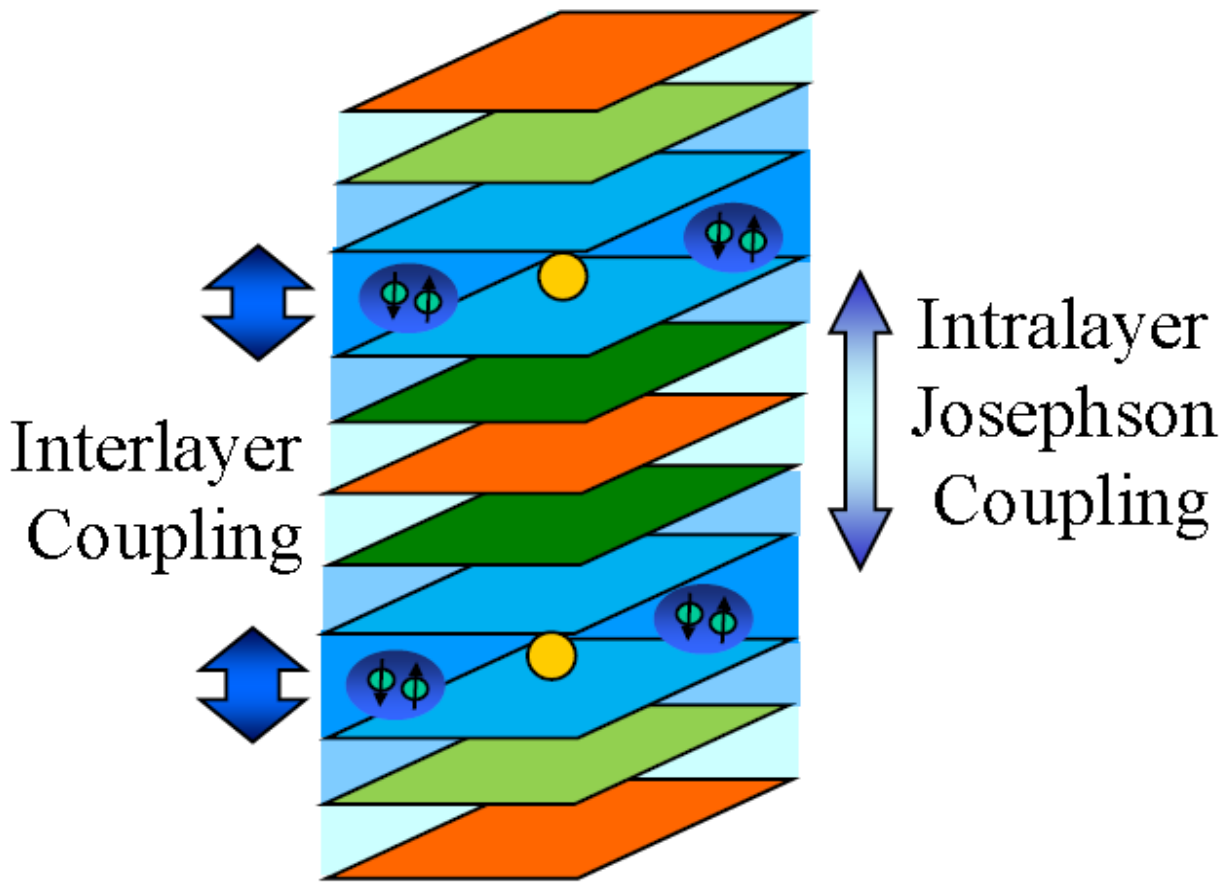

**Fig. 3** Hierarchy of the superconductive coupling in cuprate HTS.

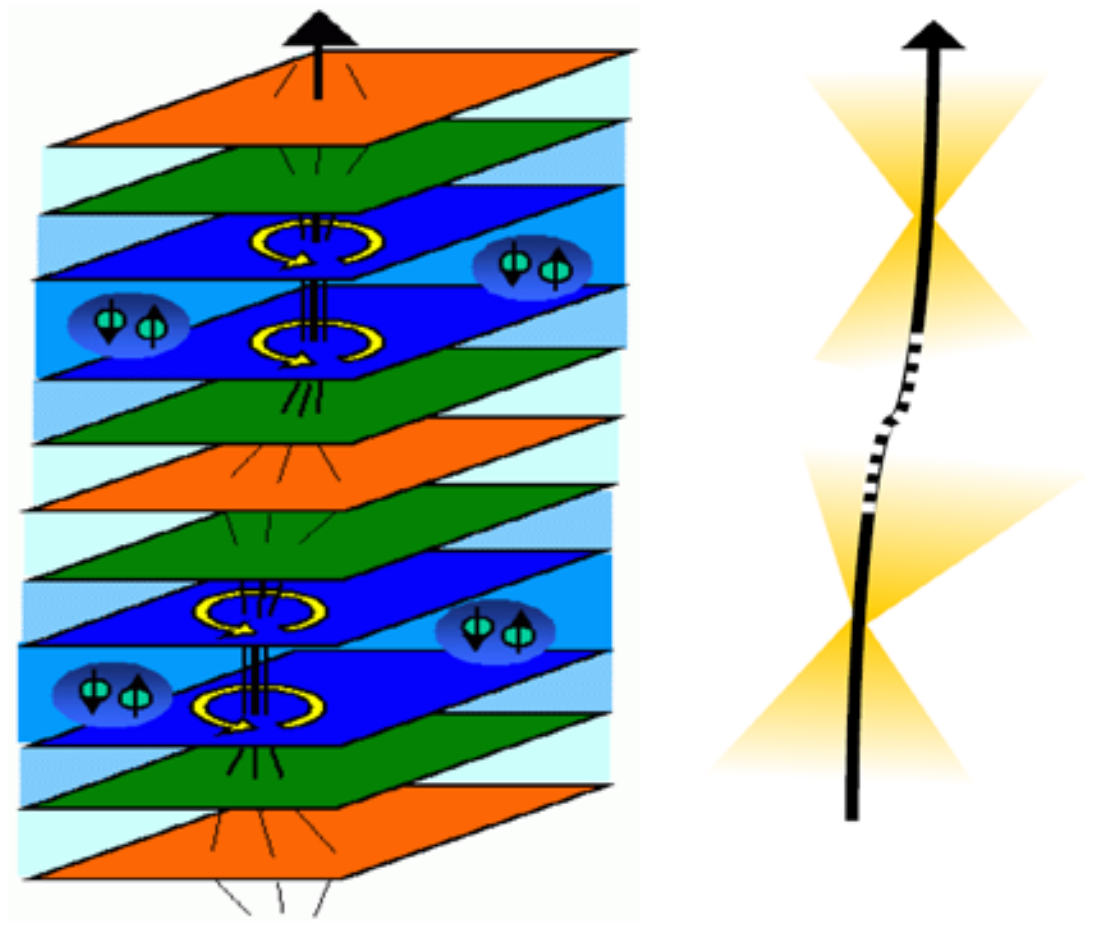

**Fig. 4** Quasi-disintegration of magnetic vortex lines into "pancake" vortices due to weak SC interlayer coupling and magnetic field overlap of neighboring vortices [83].

This explains the remarkable reduction of the superconductive anisotropy in c-axis direction compared to all other HTS families. Moreover, in contrast to the usually isotropic SC behavior within the a-b-plane, the CuO chains seem to introduce a substantially higher SC gap in b- compared to the a-direction [73]. This particular SC anisotropy renders YBCO and the RE-123 HTS exceptional among the cuprate HTS.

HTS are extreme type-II superconductors [74] with $\lambda > 100$ nm and $\xi \sim 1$ nm. The quasi-2-dimensional nature of superconductivity in HTS leads to a pronounced anisotropy of the SC properties with much higher supercurrents along the $CuO_2$ planes than in the perpendicular direction [75,76] and a corresponding anisotropy of $\lambda$, e.g. $\lambda_{ab}$ = 750 nm and $\lambda_c$ = 150 nm in YBCO [77] (the indices refer to the respective orientation of the magnetic field). Material imperfections of the dimension of the coherence length which are required as pinning centers preventing the flux flow of magnetic vortices are easily encountered in HTS due to their small coherence lengths, e. g., for optimally doped YBCO $\xi_{ab}$ = 1.6 nm, $\xi_c$ = 0.3 nm for $T \rightarrow 0$ K [78] which are already comparable to the lattice parameters (YBCO: $a$ = 0.382 nm, $b$ = 0.389 nm, $c$ = 1.167 nm [1]). The high $T_c$ in combination with the small value of coherence volume $(\xi_{ab})^2\xi_c \sim 1\ \text{nm}^3$ allows large thermally induced magnetic fluctuations in the SC phase at temperature close to $T_c$, an effect which could completely neglected in classical superconductors [79]. Moreover, since technical superconductor materials consist of a network of connected grains, already small imperfections at the grain boundaries with spatial extensions of the order of the coherence length lead to a substantial weakening of the SC connection of the grains and thus to a "weak-link behavior" of the transport properties. Obviously, this effect has to be avoided in technical conductor materials [80]. On the other hand it has also been widely exploited for the fabrication of HTS Josephson junctions [81].

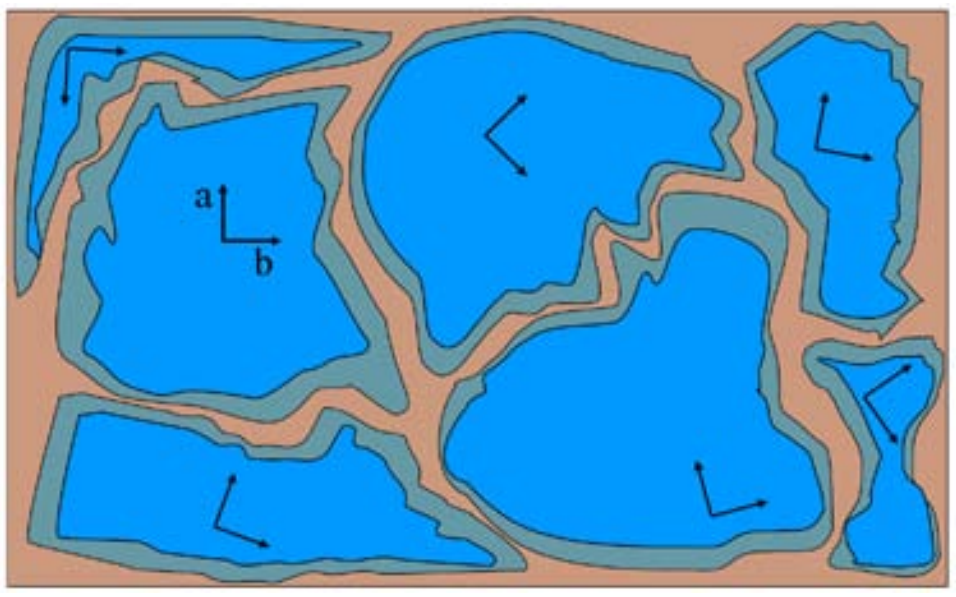

**Fig. 5** Schematics of the HTS microstructure: Differently oriented single crystal grains are separated by regions filled with secondary phase relicts from the melt growth. In addition, oxygen depletion and thus $T_c$ reduction may occur at grain boundaries.

The low $\xi_c$, i.e. the weak superconductive coupling between the $(CuO_2/Ca/)_{n-1}CuO_2$ stacks may lead for c-axis transport to an intrinsic Josephson effect within the unit cell even for perfect single-crystalline materials [82]. If the thickness of the charge reservoir blocks $EO/(AO_x)_m/EO$ in-between these stacks is larger than $\xi_c$ vortices are here no longer well-defined due to the low Cooper pair density (see Fig. 4). This leads to a quasi-disintegration of the vortices into stacks of "*pancake vortices*" which are much more flexible entities than the continuous quasi-rigid vortex lines in conventional superconductors and require therefore individual pinning centers. The extent of this quasi-disintegration is different for the various HTS compounds since $\xi_c$ is on the order of the thickness of a single oxide layers, e.g. $d_{TlOx}$ = 0.2 nm for the Tl-HTS [77]. Hence the number of layers in the charge reservoir blocks $EO/(AO_x)_m/EO$ makes a significant difference with respect to the pinning properties and thus to their supercurrents in magnetic fields. This is one of the reasons why YBCO ("Cu-**1**212") has a higher supercurrent capability in magnetic fields than the Bi-HTS Bi-**2**212 and Bi-**2**223 which for manufacturing reasons have been for a long time the most prominent HTS conductor materials. In addition, in the Cu-HTS family the $AO_x$ layer is formed by CuO chains (see Fig. 1) which apparently become SC via proximity effect. This leads here to the smallest superconductive anisotropy among all HTS families [84].

The effects described in the preceding two paragraphs combine to reduce the *irreversibility field* $B_{irr}[T]$, the tolerable limit for magnetic fields with respect to SC transport, in cuprate HTS substantially below the thermodynamical critical field $B_{c2}[T]$, a distinction which was more or less only of academic interest in the case of classical superconductor.

Beside these intrinsic obstacles for the transport of supercurrent in single-crystalline HTS materials there are additional hurdles since HTS material is not a homogeneous continuum but rather a network of linked grains (see Fig. 5). The process of crystal growth is such that all material that cannot be fitted into the lattice structure of the growing grains is pushed aside the growth front with the consequence that in the end all remnants of secondary phases and impurities are concentrated at the boundaries in between the grains. Such barriers impede the current transport even if they consist of only a few atomic layers and have to be avoided by careful control of the growth process, in particular of the composition of the offered material.

Another obstacle for supercurrents in HTS (which is detrimental for transport currents, but also enables the fabrication of HTS Josephson junctions) is misalignment of the grains: Exponential degradation of the supercurrent transport is observed as a function of the misalignment angle (see Fig. 6).

One of the reasons for this behavior is the d-symmetry of the SC order parameter (see Fig. 7) [86]: Cuprate HTS have been established as a textbook example of a d-wave symmetric SC order parameter which can be observed directly by means of particularly designed "Superconducting QUantum Interference (Device)" ("SQUID") circuits [85,86,87,88, 89,90].

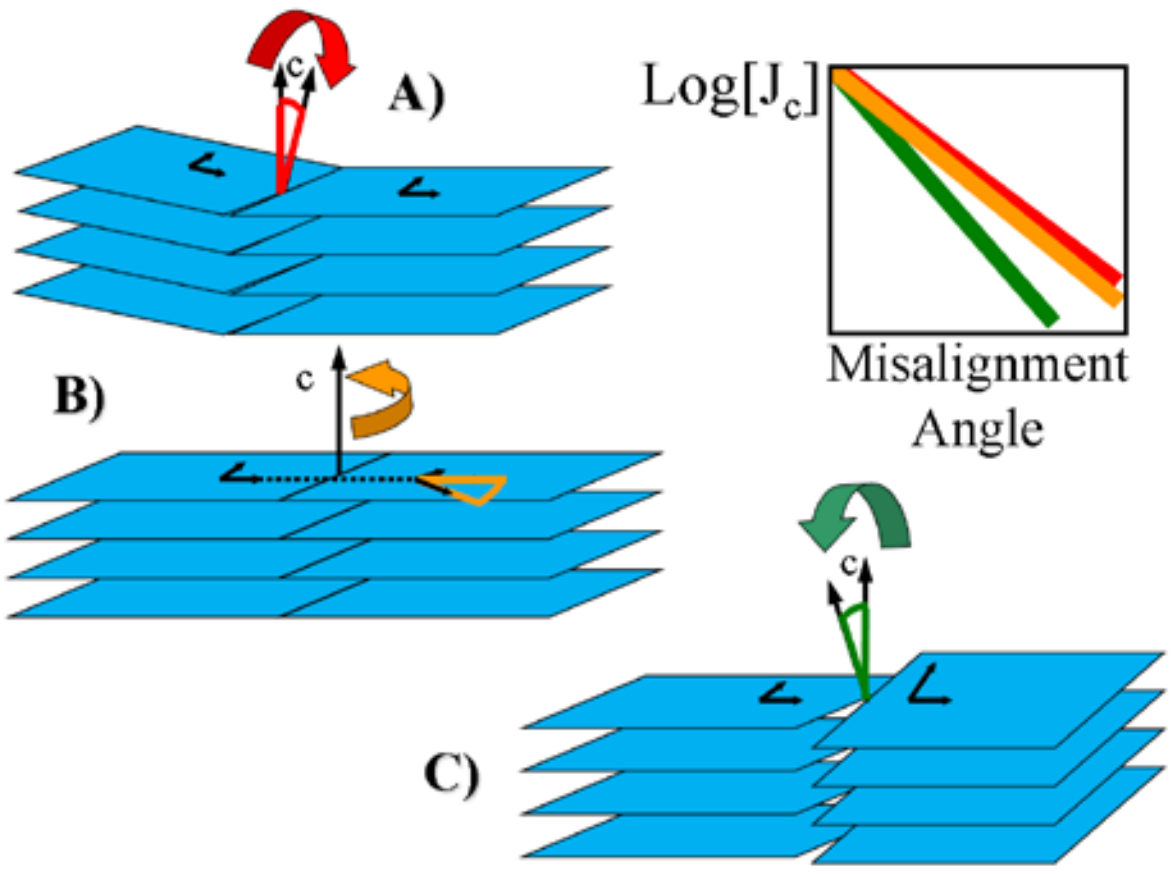

**Fig. 6** Basic grain boundary geometries and experimentally observed $J_c$ reduction $J_c \sim e^{-\alpha/\alpha 0}$ as function of the misalignment angle $\alpha$: $\alpha_0 \approx 5°$ for A) and B), $\alpha_0 \approx 3°$ for C) independent of temperature [91].

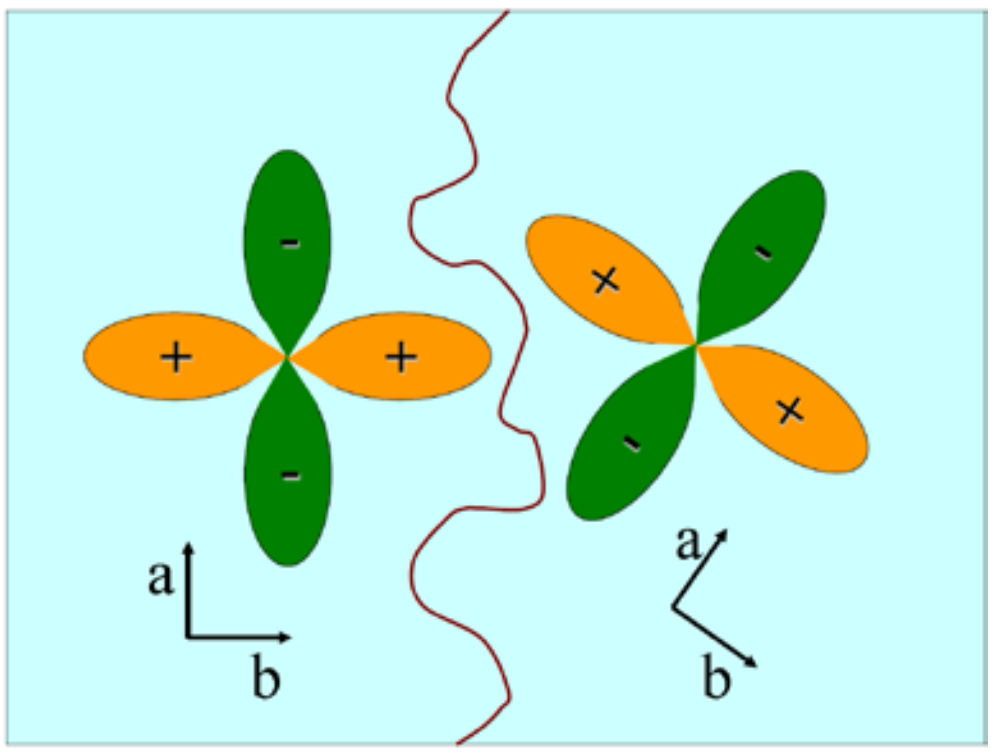

**Fig. 7** Schematics of a boundary between HTS grains. Misorientation of the SC d-wave order parameter leads to partial cancellation of the supercurrents modified by the faceting of the grain boundaries.

However, the $J_c$ reduction as a function of the misalignment angle $\alpha$ turns out to be much larger than what is expected from d-wave symmetry alone [91,92,93]. This extra $J_c$ degradation as well as the change of the current-voltage characteristics of the transport behavior [94] were believed to arise from structural defects such as dislocations [95] and deviations from stoichiometry, in particular the loss of oxygen at the grain surfaces [96] and the concomitant local degradation of the SC properties due to the decreased doping (see Fig. 2) [97]. A recent microscopic modeling identified the build-up of charge inhomogeneities as the dominant mechanism for the suppression of the supercurrent [98].

Due to all of these limitations, practical application of cuprate HTS materials has turned out to be restricted to perfectly aligned single-crystalline materials such as epitaxial films and well-textured bulk material without *weak-link behavior*, the drastic reduction of critical currents already in low magnetic fields resulting from the effects described above.

Bi-HTS *(oxide) powder in tube ("(O)PIT")* tape conductors seem to be the only exception since they constantly paved their way from the first short samples in 1989 [99] to present large cable projects [100,101]. The basic idea behind this wire preparation technique can be seen from the solution of the problem how to knot a cigarette: Wrap it in aluminum foil and then go ahead with your mechanical deformation! For Bi-HTS powder this principle works with Ag tubes as well: Bi-2212 or Bi-2223 powder or respective precursor powder is filled in Ag tubes which are subject to several mechanical deformation steps of drawing and rolling and intermediate annealing for the development of the SC Bi-HTS phase. The oxygen permeability of Ag allows for sufficient subsequent oxygenation. The two neighbouring BiO layers in the atomic Bi-2212 or Bi-2223 structure are only weakly bound and lead to graphite-like mechanical properties which allow an easy sliding or splitting of the grains along these layers (that is the chemical reason why for YBCO and Tl-1223 with only a single intermediate oxide layer OPIT wire fabrication did not meet with success). The resulting plate-like Bi-HTS grains in the filaments become aligned during the mechanical deformation steps of drawing and rolling within a few degrees. This process has been optimizied and results nowadays in conductor material with SC currents that are sufficiently high for cable applications. Unavoidably, microcracks occur in the Bi-HTS filaments, but are fortunately short-circuited by the Ag matrix which stays in close contact to Bi-HTS filaments. Nevertheless, this results in the inclusion of short resistive current paths which are reflected in the current-voltage characteristics that do not allow persistent current operation. The cost of Ag as the only possible tube material is an additional handicap. YBCO-coated metal bands with epitaxial single-crystalline YBCO grain alignment offer the prospect of overcoming both problems within foreseeable future, but are nowadays due to the required complicated fabrication process still quite expensive as well. Anyhow, even though not all cuprate HTS dreams have come true, the broad commercial interest in applications enabled by these two conductor materials will now definitively establish at least one of these cuprate HTS within the next decade as new important technical conductor material.